# From parabolic-trough to metasurface-concentrator

LiYi Hsu, Matthieu Dupré, Abdoulaye Ndao, and Boubacar Kanté[*]

Metasurfaces are promising tools towards novel designs for flat optics applications. As such their quality and tolerance to fabrication imperfections need to be evaluated with specific tools. However, most such tools rely on the geometrical optics approximation and are not straightforwardly applicable to metasurfaces. In this Letter, we introduce and evaluate, for metasurfaces, parameters such as the intercept factor and the slope error usually defined for solar concentrators in the realm of ray-optics. After proposing definitions valid in physical optics, we put forward an approach to calculate them. As examples, we design three different concentrators based on three specific unit cells and assess them numerically. The concept allows for the comparison of the efficiency of the metasurfaces, their sensitivities to fabrication imperfections and will be critical for practical systems.

## 1. Introduction

Metasurfaces, as a two-dimensional version of metamaterials, have raised significant attention due to the simplified design afforded by generalized Snell's laws of reflection and refraction [1]. They consist of arrangements of subwavelength elements and provide powerful solutions to control the phase, the amplitude and the polarization of waves at subwavelength scales. The metasurfaces can be theoretically modeled in terms of surface polarizabilities (electric and magnetic) with physical bounds [2-3]. They offer a promising platform for applications including optical devices for polarization conversion [4], beam splitters [5], beam scanning [6], carpet cloaking [7-8], holography [9] and concentrators [10-12]. Among these applications, metasurface metalens and concentrators are receiving considerable attention due to their capabilities for flat and integrable optics, super-focusing, super-imaging and solar energy.

Conventional lenses are bulky as they rely on the Snell-Descartes laws of refraction and propagation over large distances—compared to the wavelength—to focus light. On the other hand, metalenses can concentrate light with very thin surfaces—of the order of micrometers—by imposing an abrupt phase-shift to light at some interface. For instance, a parabolic metallic concentrator can be replaced by a thin and flat metasurface which provides, to a normally incoming plane wave, the parabolic phase-shift given by:

$$\Phi = k_0(\sqrt{x^2 + f^2} - f) \quad (1)$$

where $k_0$ is the free space wave-vector, x is the distance between the considered element and the center of the lens and f is the focal length. In general, only the focusing efficiency is considered to determine the quality of metalenses [11-12]. The latter corresponds to the ratio of the power incident on the focus to the power incident on the lens. In the solar concentrator field, the efficiency is defined as the ratio of solar energy collected by the receiver—an optical absorber—to that intercepted by the concentrator. The total optical efficiency of a solar concentrator is given by the combination of the so-called intercept factor, the reflectance of the concentrator, and the absorbance of the latter [14]. Since the efficiency of an energy concentrator is extremely sensitive to its geometrical parameters, it is essential to develop methods that allow their optimization.

[*] Corresponding author: B. Kanté (bkante@ucsd.edu)
The authors are with the Department of Electrical and Computer Engineering, University of California San Diego, La Jolla, California 92093-0407, USA.

In this Letter, we introduce a method to compare the quality of concentrators in the realm of metasurfaces. Specifically, we generalize the concepts of the slope error and the intercept factor. An approach based on finite difference time domain (FDTD) simulations is proposed to evaluate the efficiency of concentrators. As examples, we design in the optical domain three metasurfaces based on different unit cells (with cylindrical, rectangular and ellipsoidal elements) made of titanium dioxide ($TiO_2$) [13]. We compare the three designs with our approach and show that the rectangular element has the minimal sensitivity to fabrication imperfections.

## 2. Intercept factor and slope error

Solar concentrators are usually made of parabolic mirrors that focus light from the sun onto a receiver. The dimensions of such mirrors and their focal length can reach several meters. In such conditions, their optical properties are efficiently described within the ray-optics approximation. In the solar concentrator field, specific parameters, based on geometrical optics considerations, such as the slope error and the intercept factor are generally used to characterize the efficiency of such devices. In nanostructured dielectrics, referring to ray-optics considerations does not make sense anymore, and the traditional parameters have to be adapted within a wave optics frame. We start here by introducing the intercept factor and the slope error as discussed in the solar concentrator field. We then propose to use such concepts for metasurface concentrators. Based on these parameters we quantitatively analyze the quality of metasurface concentrators.

In the solar concentrator field [14-18], the intercept factor and the slope error allow the description of the imperfections of a solar concentrator. The intercept factor is defined as the ratio of the ray incident on the concentrator that is intercepted by the receiver. Equivalently, we can define it as the ratio of the number of rays that hit the receiver to the number of rays impinging onto the concentrator. The slope error is locally given by the difference of angle between the normal of the mirror surface with respect to the normal of the ideal surface. Ideally, the parabolic slope of the mirror is perfect, the slope error is null, all rays intercept the receiver and the intercept factor is unity. In practice, the curve of the mirror is not perfect due to fabrication imperfections, the slope of the mirror deviates from the ideal one and the slope error is finite—generally of the order of $10^{-3}$ rad. As a result, some rays are not reflected in the expected direction and miss the receiver. The intercept factor is, therefore, inferior to one, with recommended values around 95% [16].

Fig. 1(a) illustrates the slope error and the intercept factor for conventional solar concentrators. The blue parabolic curve represents the ideal mirror while the brown one represents the real mirror. A ray incident on the surface mirror is not reflected towards the initial direction due to the slope error $\Delta\theta$ but with an angular error of $2\Delta\theta$, that can make it miss the receiver if $2\Delta\theta R > D/2$, where $D$ is the diameter of the receiver and R the distance between the latter and the mirror.

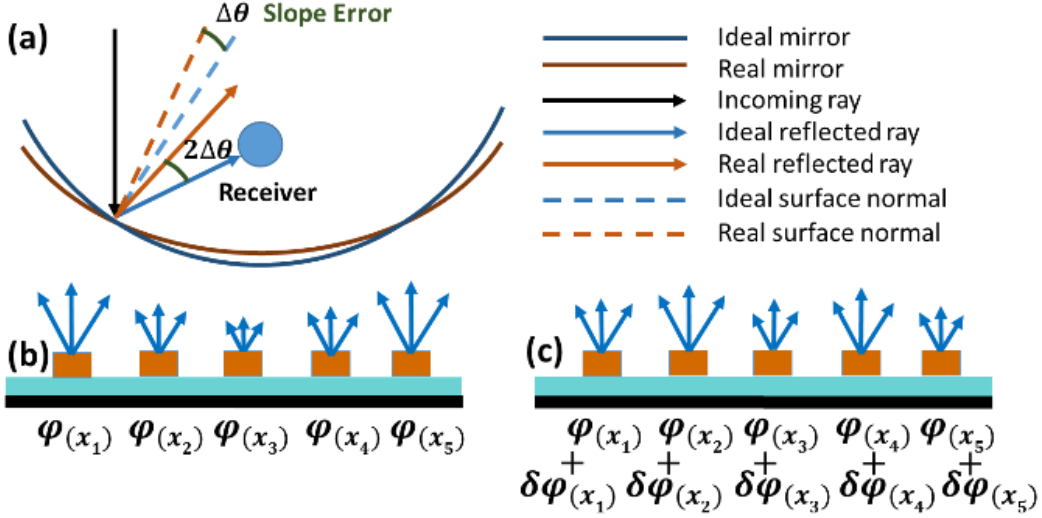

Fig. 1. (a) Schematic of the parabolic concentrator: comparison between the ideal and real cases. (b) Ideal metasurface with uniform sampling of the exact phase required for a concentrator. (c) Real metasurface with a phase variation $\delta\varphi$ due to fabrication imperfections.

In the context of metasurface lenses and concentrators, such definitions based on ray-optics cannot hold anymore. Fig. 1(b) represents an ideal metasurface which elements impose an ideal phase $\varphi$ to an incoming plane wave at different positions ($x_1, x_2,\ldots, x_n$). This is the equivalent of the ideal slope of traditional parabolic mirrors. Fig. 1(c) schemes a real metasurface which elements impose a non-ideal phase that can be modeled as the sum of the ideal phase-shift with an additional deviation $\delta\varphi(x)$ due to fabrication imperfections. With metasurface, we prefer to refer to the power rather than considering a number of rays that do not have much meaning anymore. Hence, we can re-define the intercept factor as the ratio of the integrated power on the receiver to the power incident on the metasurface. As in the ray-optics case, if there is no absorption, the intercept factor is unity for a perfect fabrication without any phase variation ($\delta\varphi(x)=0$). Generalizing the slope error is not as straightforward. In the ray-optics case, the slope error and the intercept factor are intimately related. Indeed, the loss of efficiency is characterized by the intercept factor and can be seen as the effect of the fabrication imperfections, characterized by the slope error. In other words, the slope error evaluates the cause, while the intercept factor assesses its effects. For traditional solar concentrators, the curved mirror has been used to bend light and focus it, which lead to the definition of the slope error. Metasurfaces are generally flat and rely on phase gradients and interferences to focus light. The corresponding equations for the ideal and real cases are respectively:

$$\begin{cases} \sin\theta_i - \sin\theta = \frac{1}{k_0}\frac{d\varphi(x)}{dx} \text{ for the ideal case} \\ \sin\theta_i - \sin(\theta + \delta\theta) = \frac{1}{k_0}\frac{d(\varphi(x)+\delta\varphi(x))}{dx} \text{ for the real case} \end{cases} \quad (2)$$

where $\theta_i$ is the angle of the incident plane wave, $\theta$ the angle of the reflected wave in the ideal case and $\theta + \delta\theta$ the angle of the reflected wave in the real case. This leads us to define the equivalent of the slope error, a unitless phase gradient error.

$$Slope\ Error = \left|\frac{1}{k_0}\frac{d\delta\varphi}{dx}\right| \qquad (3)$$

the cloaking mechanism, we consider two simple cases. In Fig. 1a, an incident wave is reflected by a flat ground plane. Snell's law dictates that the reflection angle is equal to the incident angle ($\theta_r = \theta_i$). In Fig. 1b, when the flat ground plane is rotated counterclockwise by an angle $\varphi$, the new incident angle becomes $\theta_i - \varphi$ while the new reflection angle becomes $\theta_r + \varphi$. Approximating each point of the Gaussian scatterer locally by a flat plane, we can design the entire cloak simply based on the geometric considerations made in Figs. 1A-B.

## 3. Quantitative analysis of Intercept factor and slope error in non-perfect metasurface

The phase shift induced by the resonant elements strongly depends on the difference between the operating frequency and the resonant frequency, and the latter is mainly sensitive to the elements dimensions. Hence, the dominant source of phase error is the fabrication imperfection resulting in mismatches between the obtained dimensions of the elements and their nominal values. Most of the time, these imperfections can be modeled as a random noise on the dimensions and positions of the elements, as any constant bias can be overcome by carefully adapting the fabrication process.

In order to show how such fabrication imperfections can degrade the efficiency of a metasurface, and how they can be characterized by the intercept factor and by the slope error, we designed metasurfaces with three different geometrical structures: cylinders (Fig. 2(a)), rectangular parallelepipeds (Fig. 2(b)), and ellipses (Fig. 2(c)) that are widely used to design elements of metasurfaces [10-12]. We chose to work in reflection and in the visible spectrum at 800 nm to illustrate the concept. Therefore, we place our elements on a metallic ground plane, in black color in Fig. 2, and which will be modeled as Perfect Electric Conductor (PEC) in our FDTD simulations. Hence, we optimize their quality factor Q by tuning the thickness of a $SiO_2$ layer—represented in blue in Fig. 2—with a refractive index of 1.45 at the considered wavelength. On top of it, we deposit our $TiO_2$ elements—represented in brown in Fig. 2—with a refractive index of 2.52 for λ=800 nm [19-20].

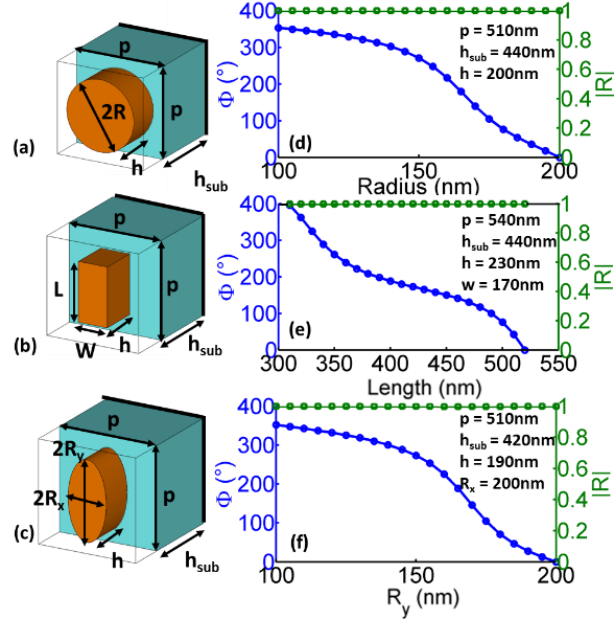

Fig. 2. (a-c) Schematic of a conventional metasurface structure (cylinders, rectangular parallelepipeds and ellipses.) with the geometrical parameters: p is the period, h is the thickness of the resonator, R is the radius, $h_{sub}$ is thickness of the substrate, W is the width, $R_x$ is the radius in x direction and $R_y$ is the radius in y direction. The brown color represents the $TiO_2$ material, the blue-green is $SiO_2$ and the black bold represents the ground plane. (d-f) Phase shift for different size of the cylinders, rectangular parallelepipeds and ellipses.

Tuning one geometrical parameter of each element—the radius of the cylinders, the length of the rectangles and the semi-axis in the y direction of the ellipses—modifies the eigenfrequency of the Mie mode of the corresponding elements. As a consequence, the phase-shift induced to an incident wave is modified. Fig. 2 (d-f) presents the resulting phase-shifts at a wavelength of 800 nm for the three considered geometries. The results are obtained with a commercial finite element code, with unit cell boundary condition in the plane of the metasurface. The cylinders and ellipses look very similar as they have closely related geometries, but they strongly differ from the rectangles, even if we could have expected the ellipsoidal elements to be an intermediary element between the cylindrical and rectangular ones. Furthermore, we can anticipate that the steepest the curve, the more sensitive is the element towards fabrication imperfections as a small parameter change will produce a large phase-shift.

In what follows, we propose to investigate more quantitatively the effects of fabrication imperfections. A solution would consist in fabricating the three metasurfaces, but it would be difficult to precisely tune the randomness of the fabrication imperfections. Therefore, we propose to adopt a simpler approach relying on FDTD simulations. We modeled the fabrication imperfections as a random phase noise that adds to the phase shift of the elements. Hence, the total phase at each element is given by, $\Phi_{real} = \Phi + \varepsilon \, \Delta P(\Phi)$. Where $\Phi_{real}$ is the phase of the element with fabrication imperfections, $\Phi$ is the ideal parabolic phase. $\varepsilon$ is a random number between -0.5 and 0.5, picked up from a uniform distribution. The type of the distribution is not really important as long as its mean and standard deviation are defined and that all elements have independent imperfections, thanks to the central limit theorem. ΔP is the magnitude of the random number which is a function of the phase shift Φ. We will now link the fabrication imperfections to the phase noise. Fig. 3 illustrates our approach for the cylindrical element. Fig. 3(a) presents the phase shift Φ(R), provided by the elements as a function of the radii. Then, we calculate |dΦ(R)/dR| the absolute value of the derivative of the phase-shift with respect to the radius—shown in Fig. 3(b)—which can be seen as a quantitative measure of the phase sensitivity to the size variation. This value is maximum for the elements which are the most sensitive to noise, i.e. the points of Fig. 3(a) for which the slope of the curve is maximum. We multiply the sensitivity by the average size variation (ΔR) (see Fig. 3(c)). In real experiments, such value is around 10 nm for conventional electron beam lithography techniques [21]. Hence, the curve plot in Fig. 3(c) represents the average phase error—in degree—for an element with radius R. Finally, we convert the parameter size axis into its phase shift (Φ) counterpart, using the bijection of Fig. 3(a). This curve, presented in Fig. 3(d) shows the average phase noise on an element for the desired phase Φ. The maximum of this noise reaches 80° for a phase shift of 170° and which corresponds to a radius R equals to 170 nm.

Using this approach, we calculate the sensitivity of the phase-shift for the rectangles and the ellipses. Results are presented in Fig. 4. As noticed before, the sensitivity of the cylinders and ellipses are similar with a maximum around 180 degrees, while the rectangular elements have a minimum of sensitivity around this value. We only considered variations in one dimension, even though, in practical fabrication, elements could have variations in other dimensions which would reduce the intercept factor. However, our approach is general and can be expanded to consider more fabrication imperfections.

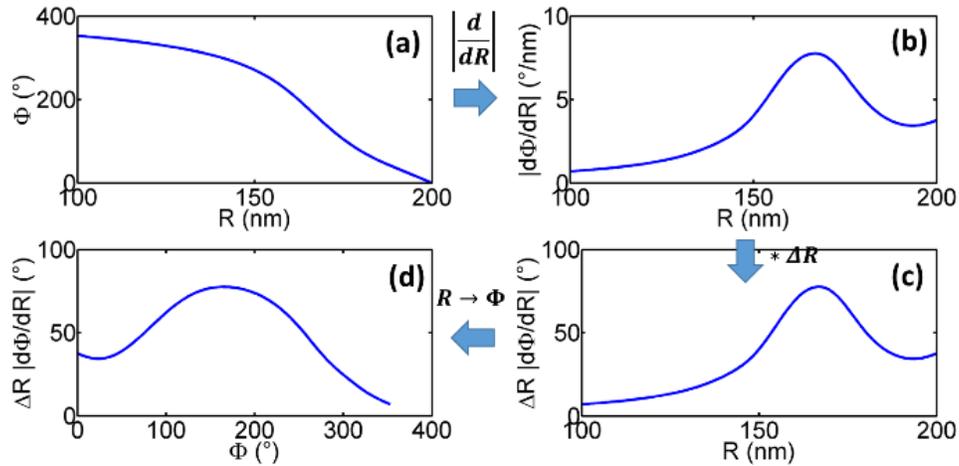

Fig. 3. Flowchart of transformation. (a) The phase shifts as a function of a geometrical parameter. (b) The sensitivity of the phase shifts to the parameter resolution in degree/nm. (c) Same in absolute value for an average noise of 10 nm. (d) Sensitivity in degree as a function of the targeted phase shift.

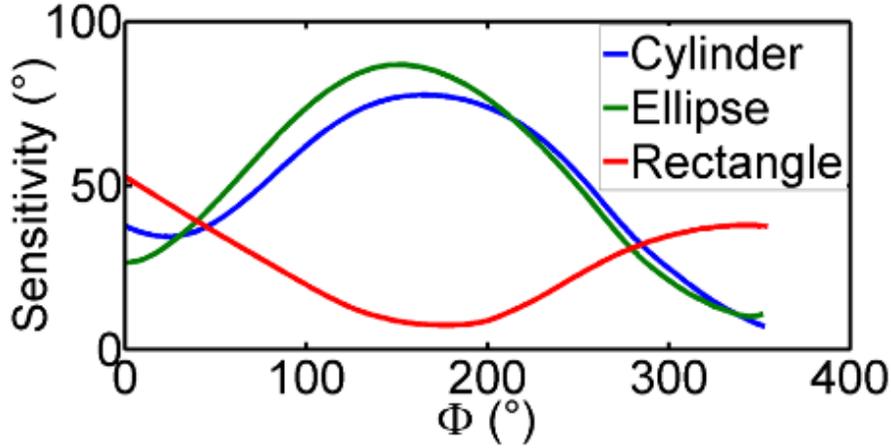
Fig. 4. Sensitivity for cylinders, ellipses and rectangular parallelepipeds.

To statistically analyze our metasurface concentrators, we run 100 simulations for each element using a homemade FTTD code. Each simulation was given a certain magnitude of the noise related to the type fabrication imperfections. We simulate a 115μm-long metasurface with a focal length of 170μm, and 241 elements. To reduce the computational time, we replace the wave reflected by the metasurface elements by point sources. Their phases are set to match the phase of a concentrator with a random spatial phase noise given by the approach described in Fig. 3 and a given magnitude of the fabrications imperfections (from 0 to 20 nm). Based on this method, we calculate the intercept factors of metasurfaces designed with the three common elements.

Fig. 5($A_1$) shows the intercept factor as a function of fabrication imperfections. For a structure without fabrication imperfections, the value of the intercept factor is equal to unity as illustrated in Fig. 5($A_1$). The field is very well focused (see Fig. 5($B_1$)). For example, for a fabrication imperfection value equal to 10 nm for cylinders and ellipses, the value of the intercept factor is about 0.72 (as illustrate in Fig. 5($B_2$)). In the case of the rectangular parallelepipeds a similar intercept factor of 0.71 (see Fig. 5($B_3$)) is obtained for twice as larger fabrication imperfections (20nm). For this same value of fabrication imperfections in the case of a cylinder and the ellipse, the value of the intercept factor is 3 times smaller than the rectangular one. (Fig. 5($B_4$)). Fig. 5($A_2$) presents the slope error as a function of the fabrication imperfections. The ellipses and the cylinders metasurface have the approximately equal slope error value that is about twice that of the rectangle. This proves that the rectangular parallelepipeds are less sensitive to the considered fabrication imperfections, and they would be more advantageous to use to design highly efficient metasurfaces.

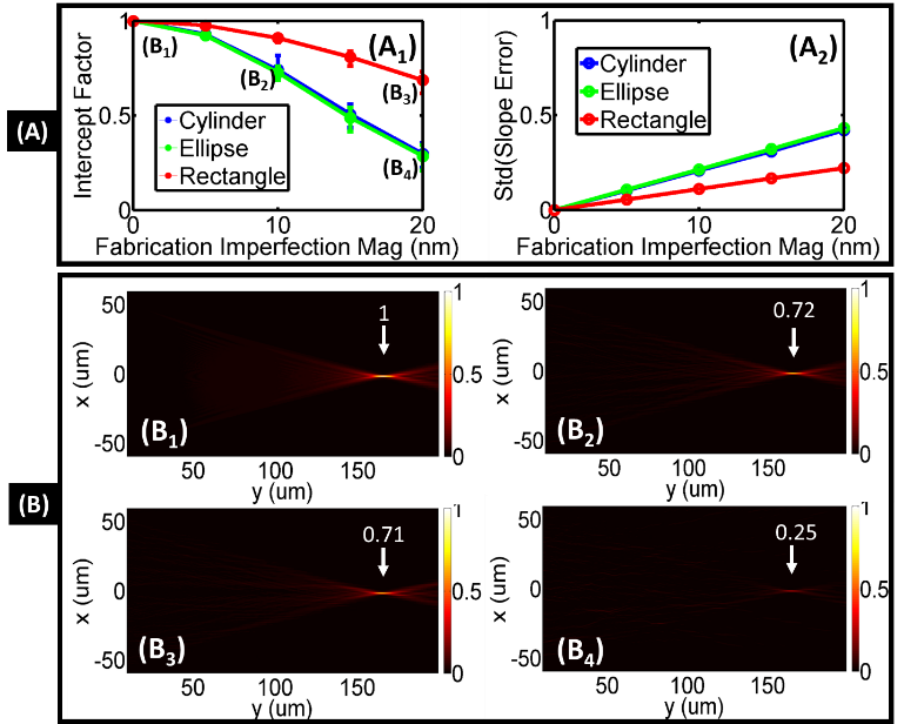

Fig. 5. ($A_1$) Intercept factor and ($A_2$) Standard variation of the slope error for cylinders, rectangular parallelepipeds and ellipses. (B) Power pattern.

## 4. Conclusion

In conclusion, we presented an approach to evaluate the robustness of metasurface concentrators to fabrication imperfections. We started by describing the general methods used, and investigated three different geometries as unit cell elements with cylinders, rectangular parallelepipeds, and ellipses cross sections. We studied the effects of imperfection via the intercept factor and the slope error. Specifically, we have computed these quantities and shown that the rectangular parallelepipeds are less sensitive to fabrication imperfections compared to the ellipse and the cylinder. Our approach can provide a guidance to design large scale and highly efficiency metasurface concentrators.